\documentclass[conference]{IEEEtran}
\IEEEoverridecommandlockouts

\usepackage{cite}
\usepackage{amsmath,amssymb,amsfonts}
\usepackage{algorithmic}
\usepackage{graphicx}
\usepackage{textcomp}
\usepackage{xcolor}
\def\BibTeX{{\rm B\kern-.05em{\sc i\kern-.025em b}\kern-.08em
    T\kern-.1667em\lower.7ex\hbox{E}\kern-.125emX}}

\usepackage{bm} 
\usepackage{physics} 
\usepackage{qcircuit} 
\usepackage{soul} 

\DeclareMathAlphabet{\bbold}{U}{bbold}{m}{n}
\newcommand{\id}{\ensuremath{\bbold{1}}}  

\begin{document}

\title{Non-variational Quantum Combinatorial Optimisation}

\author{\IEEEauthorblockN{1\textsuperscript{st} Tavis Bennett}
\IEEEauthorblockA{\textit{Department of Physics} \\
\textit{The University of Western Australia}\\
Perth, Australia \\
tavis.bennett@research.uwa.edu.au}
\and
\IEEEauthorblockN{2\textsuperscript{nd} Lyle Noakes}
\IEEEauthorblockA{\textit{Department of Mathematics and Statistics} \\
\textit{The University of Western Australia}\\
Perth, Australia \\
lyle.noakes@uwa.edu.au}
\and
\IEEEauthorblockN{3\textsuperscript{rd} Jingbo Wang}
\IEEEauthorblockA{\textit{Department of Physics} \\
\textit{The University of Western Australia}\\
Perth, Australia \\
jingbo.wang@uwa.edu.au}
}

\maketitle

\begin{abstract}
This paper introduces a non-variational quantum algorithm designed to solve a wide range of combinatorial optimisation problems, including constrained and non-binary problems. The algorithm leverages an engineered interference process achieved through repeated application of two unitaries; one inducing phase-shifts dependent on objective function values, and the other mixing phase-shifted probability amplitudes via a continuous-time quantum walk (CTQW) on a problem-specific graph. The algorithm's versatility is demonstrated through its application to various problems, namely those for which solutions are characterised by either a vector of binary variables, a vector of non-binary integer variables, or permutations (a vector of integer variables without repetition). An efficient quantum circuit implementation of the CTQW for each of these problem types is also discussed. A penalty function approach for constrained problems is also introduced, including a method for optimising the penalty function. The algorithm's performance is demonstrated through numerical simulation for randomly generated instances of the following problems (and problem sizes): weighted maxcut (18 vertices), maximum independent set (18 vertices), k-means clustering (12 datapoints, 3 clusters), capacitated facility location (12 customers, 3 facility locations), and the quadratic assignment problem (9 locations). For each problem instance, the algorithm finds a globally optimal solution with a small number of iterations.
\end{abstract}


\section{Introduction}

In 1996 Grover introduced his landmark quantum search algorithm, addressing the problem of unstructured search \cite{grover_search}. Grover's algorithm leverages quantum superposition, entanglement and interference to return a target element from an unstructured search space with a quadratic speedup relative to exhaustive search ($\mathcal{O}(\sqrt{N})$ vs. $\mathcal{O}(N)$). Since then, the amplitude amplification process at the heart of Grover's algorithm has formed an important sub-process within many quantum algorithms, delivering a similar quadratic speedup \cite{brassard1998quantum,suzuki2020amplitude,brassard1997exact,montanaro2015quantum}. 

Soon after, amplitude amplification was also applied to combinatorial optimisation. The Grover adaptive search treats the space of feasible solutions to a combinatorial optimisation problem as an unstructured database, where solutions are differentiated from each other only by their associated objective function values \cite{GAS,GAS2}. However, with only a quadratic speedup relative to an exhaustive search, this technique does not yield a significant advantage over classical techniques which instead take advantage of problem structures. 

A more promising candidate for combinatorial optimisation, the quantum approximate optimisation algorithm (QAOA) \cite{Farhi2014QAOA}, inspired by quantum annealing and the adiabatic theorem \cite{Adiabatic_QC,kadowaki1998quantum,Adiabatic_theorem}, makes use of a variational approach. The algorithm sees the repeated application of two Hamiltonians where the application times are classically controlled parameters. These parameters would typically be initialised randomly, and subsequently tuned via a classical optimisation process. The QAOA is not the only variational algorithm for solving combinatorial optimisation problems; the variational quantum eigensolver and related algorithms \cite{peruzzo2014variational,amaro2022filtering} involve the application of parameterised quantum circuits for which the parameters must be similarly tuned. A major challenge with these variational algorithms is finding a set of parameters that performs well and returns an optimal or near-optimal solution. While gradient based methods can be used to improve a set of parameters, the resulting performance is highly sensitive to initialised parameter values. With the space of possible initial parameter values growing exponentially in parameter count, without some method of generating an effective set of initial parameters, variational algorithms aim to solve a hard problem, while presenting another in its place.

Besides the challenge of producing a good set of parameters, without a clear understanding of the mechanism by which variational quantum circuits increase the measurement probability of optimal solutions, it is also not clear how to design or select an effective circuit ansatz. For example, the QAOA relies on parameterized applications of a particular (transverse-field) Hamiltonian, yet there exist many other efficiently implementable Hamiltonians that could potentially be used in its place. While the transverse-field Hamiltonian is effective with binary variable problems, such as maxcut, it does not perform as well in non-binary problems. In the absence of a more adaptable approach, many studies opt to embed intrinsically non-binary problems into a quadratic unconstrained binary optimisation (QUBO) framework, introducing additional problem constraints (via a penalised objective function) to enforce the original problem structure \cite{Azad_QUBO_VRP,papalitsas2019qubo}. 

The quantum walk-based optimisation algorithm (QWOA) is another variational algorithm which is particularly relevant to this discussion, due to its close connection with the non-variational algorithm of this paper. The QWOA is a generalisation of the QAOA, where the application of the transverse-field Hamiltonian is replaced by a continuous-time quantum walk (CTQW)~\cite{QWbook2014} on a graph that connects basis states encoding the feasible solutions to a combinatorial optimisation problem \cite{marsh2019quantum,marsh2020combinatorial,slate2021quantum,bennett2021quantum}. The other Hamiltonian remains unchanged, acting to phase-shift basis states depending on their associated objective function values. Within this framework, QAOA's transverse-field Hamiltonian is equivalent to a CTQW on a hypercube, connecting computational basis states which differ by a single bit-flip.

Grover's algorithm can also be described within the QWOA framework: Grover's diffusion transform is equivalent to a CTQW on a complete graph for time $\frac{\pi}{N}$, while Grover's other unitary is implementable with $\pi$ phase-shifts applied with a binary objective function. The complete graph CTQW distributes phase-shifted probability amplitudes across the connected basis states, producing a well-designed interference process, where probability amplitudes arriving at target basis states constructively interfere, amplifying the associated measurement probability. Grover's algorithm cannot exploit problem structure as the complete graph connects any particular basis state to all others, ignoring any relationship that may be present between them. The binary phase rotation also erases much of the structure that may have been present across an optimisation landscape consisting of various objective function values. 

On the other hand, performing the CTQW over a graph that incorporates the underlying problem structure enables exploitation of this structure, leading to speedups significantly larger than that of Grover's algorithm. This paper introduces an algorithm that does exactly that, the non-variational quantum walk-based optimisation algorithm (non-variational QWOA). By closely analysing the CTQW and studying the statistics of objective function values as distributed over the mixing graph, it is possible to understand and design the algorithm's interference process. This understanding helps with the intelligent selection of parameters, namely those that control the walk times and magnitudes of applied phase-shifts, thus removing the need for a computationally expensive variational procedure. In addition, a clear understanding of the interference process informs the design of the mixing graph, allowing for generalisation to problems with diverse and non-binary structures. Effective penalty functions can also be designed to embed problem constraints where necessary, further improving versatility.

This paper formally introduces the non-variational QWOA and focuses on its application to various combinatorial optimisation problems. Efficient circuit implementations of CTQWs for binary, non-binary and permutation problems are discussed, after which the algorithm's performance is analysed via numerical simulation results. Detailed benchmarking will form part of future work and a more detailed analysis of the algorithm, particularly the general interference process, and the problem specific CTQWs is included in a separate paper \cite{longPaper}.

\section{The Algorithm (non-variational qwoa)}

\subsection{Definitions}
Consider a combinatorial optimisation problem with a solution space $S$ containing $N$ feasible solutions ${\bm{x} = (x_1, x_2, ..., x_n)}$ each composed of $n$ decision variables, ${x_j \in \{0,1,...,k-1\}}$. The solution space may contain all possible solutions of this form, of which there are $N=k^n$. Alternatively, it may be restricted to solutions which satisfy some constraint, such as in the case of permutation-based problems where $k=n$ and $N=n!$. The solutions to the problem can be encoded in the computational basis states of a quantum computer by allocating to each variable a sub-register such that $\bm{x}$ is represented by the solution state,
\begin{equation}
    \ket{\bm{x}} = \prod_{j=1}^n \ket{x_j},
\end{equation}
where $\ket{x_j}$ is the computational basis state of the $j^{\text{th}}$ sub-register which directly encodes the decision variable $x_j$. If using a binary encoding, each sub-register will be assigned $\lceil \log_2 k \rceil$ qubits, whereas a one-hot encoding will use $k$ qubits. The binary encoding is more space efficient and lends itself to a more efficient implementation of the mixing unitary. However, the one-hot encoding may enable a more efficient implementation of the phase-separation unitary.

The state of the allocated qubits is initialised in the equal superposition state, with equal probability amplitude assigned to each solution state,
\begin{equation}
    \ket{s} = \frac{1}{\sqrt{N}} \sum_{\bm{x} \in S} \ket{\bm{x}}.
\end{equation}
The preparation of this equal superposition state is discussed in Section \ref{sec:mixers}.

The non-variational QWOA is designed to find optimal or near-optimal solutions to the combinatorial optimisation problem, with respect to optimisation of an objective function $f(\bm{x})$. The interference process responsible for amplifying the measurement probability of optimal and near-optimal solution states is driven by repeated applications of two unitary operations. The first of these is the phase-separation unitary,
\begin{equation}
    U_Q(\gamma) = e^{-\text{i} \gamma Q},
    \label{eq:U_Q}
\end{equation}
where $Q$ is a diagonal operator such that ${Q\ket{\bm{x}} = f(\bm{x})\ket{\bm{x}}}$. This unitary applies a phase shift to each solution state, proportional to the associated objective function value,
\begin{equation}
    U_Q(\gamma)\ket{\bm{x}} = e^{-\text{i} \gamma f(\bm{x})} \ket{\bm{x}}.
\end{equation}

The second unitary performs a continuous-time quantum walk for time $t$ on the mixing graph which connects feasible solution states, 
\begin{equation}
    U_M(t) =  e^{-\text{i} t A},    
\end{equation}
where $A$ is the adjacency matrix which defines the mixing graph's structure.  This unitary is referred to as the mixing unitary, or mixer for short, since it acts to distribute probability amplitudes between the solution states. The mixing graph is customised to suit the underlying structure of the problem to be solved. The efficient implementation of this mixer is discussed in Section \ref{sec:mixers}.

\subsection{The Mixing Graph}
The adjacency matrix $A$ is formed by connecting solution states associated with solutions that are nearest neighbours. For the mixing graphs presented in this paper, nearest neighbour solutions are defined as those separated by a minimum non-zero Hamming distance, where the Hamming distance between any two solutions counts the number of mismatched decision variables between them. This generally corresponds with a Hamming distance of $1$, though not necessarily, as in the case of permutations, where the smallest possible Hamming distance between two different permutations is $2$.

More generally, we specify a set of $d$ moves (polynomial in problem size), each of which modifies any feasible solution to return a different feasible solution with similar configuration. Given a particular solution, its nearest neighbours are those that can be generated with a single move from the set of available moves. Distance on the mixing graph therefore acts as a measure of similarity, with the distance between any two vertices counting the minimum number of moves required to transform between the associated solutions. In addition, since the nearest neighbours of each solution are generated by the same set of $d$ moves, the graph is vertex transitive with degree $d$.

\begin{figure}[htbp]
    \centering
    \includegraphics[width=1\columnwidth]{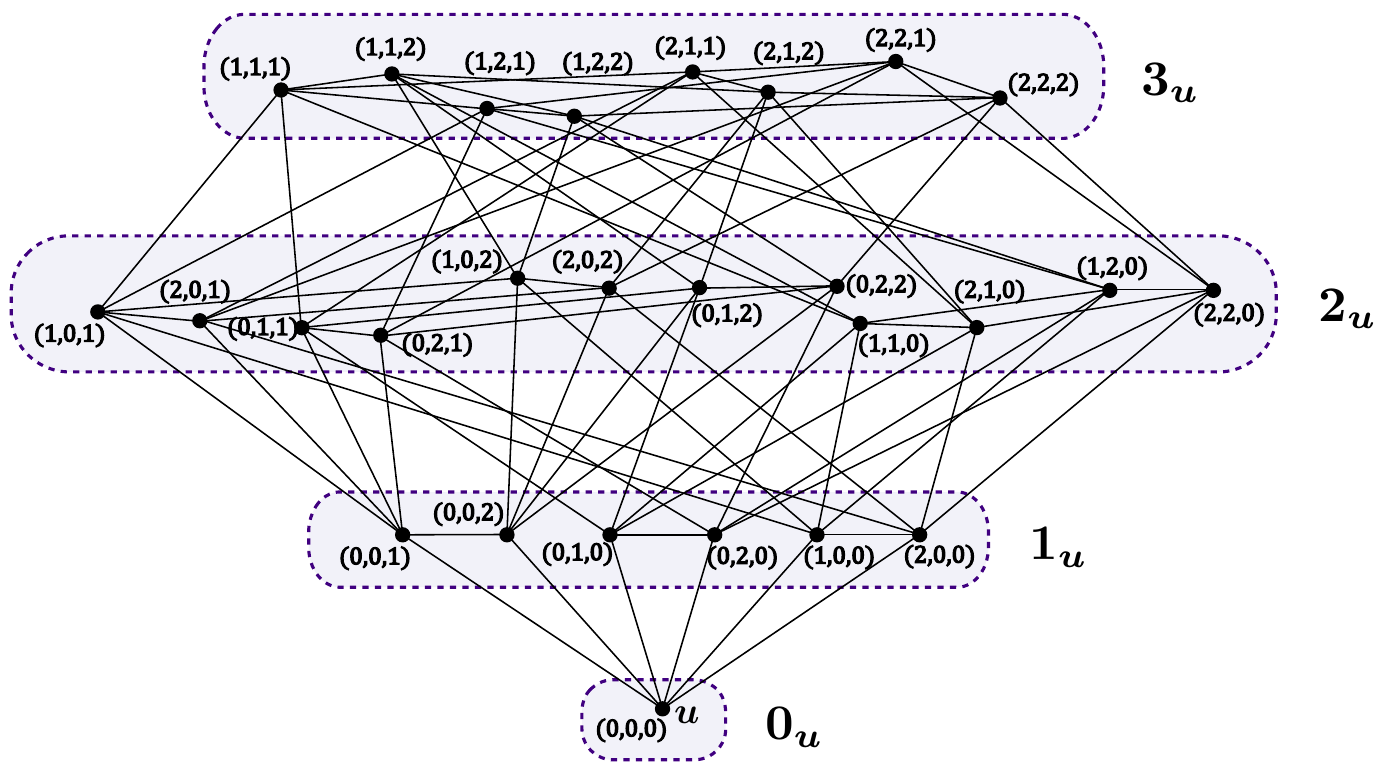}
    \vspace{-0.2cm}
    \caption{Example mixing graph connecting solutions to an integer variable problem, for $n=3$, $k=3$. Solution $\bm{u}=(0,0,0)$ is selected, and the graph is partitioned into subsets $h_{\bm{u}}$ for $h=0,1,2,3$.}
    \label{fig:vertex_subsets}
\end{figure}

The mixing graph defined by the adjacency matrix should have a diameter $D$ which scales linearly in the size of the problem instance. For effective performance the mixer should also satisfy two important conditions, but before describing these, it is useful to first define distance-based subsets of vertices on the graph. For any two vertices, $\bm{u}$ and $\bm{v}$, we define ${\text{dist}(\bm{u},\bm{v})}$ as the distance on the graph between these vertices. For any vertex $\bm{u}$ and distance $h$, we define a subset of vertices,
\begin{equation}
    h_{\bm{u}} = \{\bm{v} : \text{dist}(\bm{u},\bm{v}) = h\},
\end{equation}
i.e. $h_{\bm{u}}$ is the subset of vertices which are a distance $h$ from $\bm{u}$ on the graph.  Fig.~\ref{fig:vertex_subsets} provides an illustration of the partitioning of a mixing graph into these distance based subsets for a particular choice of vertex $\bm{u}$.

It may often be the case where the mixing unitary is applied over a quantum register with a Hilbert space larger than the feasible solution space (as in the mixer corresponding with Fig.~\ref{fig:vertex_subsets} where the Hilbert space of each sub-register has dimension larger than $k=3$). In this case, the adjacency matrix defines a disconnected graph with several components (isolated subgraphs). The relevant subgraph connecting just the feasible solution states, which we refer to as the mixing graph, is just one of the components, where the others can be ignored. This is because the continuous-time quantum walk distributes probability amplitude only between connected vertices/states, such that the feasible solution states form an invariant subspace of the mixer.

\subsection{Necessary Conditions}
Here we impose two conditions under which the algorithm is expected to work well. These conditions are satisfied within a surprisingly wide range of problems, as demonstrated in Sections \ref{sec:mixers} and \ref{sec:results}. The first condition, which relates to selecting a neighbourhood and designing a mixing graph, is that the action of the resulting mixer on an arbitrary solution state $\ket{\bm{u}}$, must be as follows:
\begin{equation}
    U_M(t) \ket{\bm{u}} = \sum_{h=0}^D \left( e^{-\text{i} h \phi(t)} \sum_{\bm{x} \in h_{\bm{u}}} r_{\bm{x}}(t) \ket{\bm{x}} \right),
    \label{eq:mixer_requirement_1}
\end{equation}
where $D$ is the diameter of the graph and (for sufficiently small $t$) $r_{\bm{x}}$ and $\phi$ are positive real-valued functions, also $0 < \phi(t) < \pi$. In other words, probability amplitude from an initial vertex is distributed to other vertices such that the complex phase of the distributed probability amplitudes is proportional to their distance from the initial vertex.

The second condition relates to the tendency for solutions with similar configurations to possess similar objective function values, and manifests due to the mixing graph clustering solutions with similar configurations. Firstly, the following relationship should be at least approximately satisfied,
\begin{equation}
    \left( \mu_{h\bm{x}} - f(\bm{x}) \right) \approx - \alpha_h \left( f(\bm{x}) - \mu \right),
    \label{eq:mixer_requirement_2}
\end{equation}
where $\mu_{h\bm{x}}$ is defined as the mean objective function value of solutions contained in $h_{\bm{x}}$, $\mu$ is the mean objective function value of all solutions in $S$, and $\alpha_h$ is a positive constant. Secondly, the constant of proportionality $\alpha_h$ should increase monotonically with increasing $h$ (up to a distance which is a considerable fraction of the graph's diameter $D$). Examples of this relationship are plotted in section \ref{sec:results} (Fig.~\ref{fig:maxcut_subset_means},~\ref{fig:MIS_subset_means},~\ref{fig:kmeans_subset_means},~\ref{fig:CFLP_subset_means},~\ref{fig:QAP_subset_means}). 

\subsection{The Amplified State}
Having established the phase-separation and mixing unitaries, and given a pre-selected number $p$ of iterations (which should be polynomial in the size of the problem instance), we define the amplified state,
\begin{equation}
    \ket{\gamma,t,\beta} = \left[ \prod_{i=0}^{p-1} U_M\left(t_i\right) U_Q\left(\frac{\pm \gamma_i}{\sigma}\right) \right] \ket{s},
    \label{eq:Amplified_state}
\end{equation}
where $\gamma$, $t$ and $\beta$ are classically controlled positive-valued parameters that determine the applied phase separations and the mixing times for each iteration. Specifically, the values of $\gamma_i$ and $t_i$ are given by explicit formulae, such that $\gamma_i$ increase linearly over the domain $[\beta\gamma,\gamma]$, while the mixing times $t_i$ decrease linearly over the domain $[\beta t, t]$, where $0 < \beta < 1$. These increasing and decreasing profiles are reminiscent of the annealing protocols for the adiabatic algorithm, though they are independently motivated. Here, $\pm$ accounts for whether the goal is maximisation ($+$) or minimisation ($-$), and $\sigma$ is the standard deviation of objective function values, controlling for variation across problems and problem instances such that appropriate values of $\gamma$ are consistently of order $\gamma \approx 1$. An approximate value for $\sigma$ is sufficient (acquired through random sampling, for instance). 

\subsection{Repeated State Preparation and Measurement}

Subject to the conditions described above, and allowing a sufficient number of iterations (polynomial in problem size), a wide range of appropriate values for $\gamma$, $t$ and $\beta$ significantly increase the measurement probability of globally optimal and near-optimal solutions within the amplified state $\ket{\gamma,t,\beta}$, such that a globally optimal or near-optimal solution is exceedingly likely to be measured following repeated state preparation and measurement of $\ket{\gamma,t,\beta}$.  

The total number of state preparations should be fixed, regardless of iteration count $p$ or problem size. 
The best solution measured during the process of repeated state preparation and measurement is taken as the solution, either exact or approximate, to the optimisation problem. Optionally, at regular intervals during the repeated state preparation and measurement, the values of $\gamma$, $t$ and $\beta$ can be updated in order to improve the amplification of optimal and near-optimal solutions, as indicated by improvement in approximated values of either the expectation value of the objective function, or a related measure, such as the Conditional Value at Risk (CVaR) \cite{barkoutsos2020improving}. 

The reason that the algorithm remains effective while constrained to a fixed total number of state-preparations and measurements, and the reason it is best characterised as non-variational, is because the optimal set of parameters can be determined via a fixed complexity 3-dimensional optimisation which seeks only a local extremum (that closest to the origin). In any case, even sub-optimal parameter values significantly amplify optimal and near-optimal solutions, such that the improvement/optimisation of parameters is optional.

\subsection{Constrained Problems}

There are two ways that problem constraints can be dealt with in this framework. The first approach is to restrict the space of feasible solutions $S$ to include just those solutions which are valid (do not violate the constraints). This approach is only viable if the space of valid solutions can be characterised independently of the problem instance, if it is possible to efficiently implement a mixer that connects only valid solutions and if the mixer satisfies the necessary condition in \eqref{eq:mixer_requirement_1}. An example of this approach is demonstrated in \ref{sec:QAP} for the quadratic assignment problem.

The other approach is to include both valid and invalid solutions and for the constraints to manifest as additional terms in the objective function, which act to penalise invalid solutions. This approach is often necessary, as it is frequently the case that the space of valid solutions cannot be predetermined, in which case it is not possible to design a mixing graph which isolates and mixes between only valid solutions. 

We introduce three kinds of penalty terms, the first two are designed to ensure that invalid solutions are sufficiently penalised, so that the penalised objective function is optimised by valid solutions. The first of these is a variable term which scales with the extent to which a constraint has been violated. The second is a fixed term which ensures that solutions which violate a constraint by only a small amount remain sufficiently penalised. The introduction of these first two penalty terms may adversely effect algorithm performance, due to the introduction of bimodality in the distribution of objective function values, which adversely effects conformance with condition \eqref{eq:mixer_requirement_2}. The third penalty term is introduced to correct for this bimodality and improve algorithm performance, while leaving near-optimal invalid solutions adequately penalised.

The influence of each penalty term is controlled by a positive-valued parameter, such that the penalty can be adjusted to improve performance. The penalised objective function is therefore expressed as $f(\bm{x})_{\bm{\lambda}}$, where ${\bm{\lambda} = (\lambda_1, \lambda_2, ...)}$ are the coefficients for each of the terms in the penalty function. Similar to the $\gamma$, $t$ and $\beta$ parameters, appropriate penalty parameters $\bm{\lambda}$ may be known from prior experience. In any case, the penalty parameters can be tuned during the repeated state preparation and measurement process. The general approach is summarised as follows:
\begin{enumerate}
    \item Design the penalty terms and select fixed penalty parameters $\bm{\lambda}_F$ which adequately penalise the invalid solutions, such that the globally optimal values of $f(\bm{x})_{\bm{\lambda}_F}$ belong to valid solutions. 
    \item Using the same penalty terms, define a second objective function, parameterised by tunable penalty parameters $\bm{\lambda}_T$. This objective function $f(\bm{x})_{\bm{\lambda}_T}$ is used within the phase-separation unitary.
    \item Through repeated state preparation and measurement of the amplified state, use a classical optimistion procedure to tune the parameters $\{\gamma,t,\beta,\bm{\lambda}_T\}$ (initialised with $\bm{\lambda}_T=\bm{\lambda}_F$) so as to optimise the expectation value of $f(\bm{x})_{\bm{\lambda}_F}$. In other words, optimise $_{\bm{\lambda}_T}\!\bra{\gamma,t,\beta} \hat{f}_{\bm{\lambda}_F} \ket{\gamma,t,\beta}_{\bm{\lambda}_T}$, where $\ket{\gamma,t,\beta}_{\bm{\lambda}_T}$ is the amplified state prepared using $f(\bm{x})_{\bm{\lambda}_T}$.
\end{enumerate}

\subsection{The Interference Process}
The interference process can be understood loosely by referring to the two necessary conditions \eqref{eq:mixer_requirement_1} and \eqref{eq:mixer_requirement_2}. When the mixing unitary distributes probability amplitude from one solution state to another, it induces a phase-shift proportional to the distance between the respective vertices on the mixing graph. Due to the nature of the CTQW, for a sufficient walk time, any particular solution state $\ket{\bm{u}}$ receives probability amplitudes from all other solution states. This exponentially large number of individually contributing probability amplitudes is capable of producing significant constructive interference. However, under the action of the mixer alone, the probability amplitudes largely destructively interfere, due to the distance dependent phase rotations applied to the contributions from each subset $h_{\bm{u}}$. 

Application of the phase-separation unitary rotates the phases of all contributions dependent on their associated objective function values. Due to the monotonically varying mean objective function values in each subset $h_{\bm{u}}$, the phase-separation unitary rotates the resultant probability amplitude from each subset differently, dependent on distance $h$. The distance-dependent relative phases induced by the phase-separation unitary are able to offset those induced by the mixing unitary, so as to bring the contributing probability amplitudes more or less in-phase, producing constructive or destructive interference. A closer analysis in \cite{longPaper} reveals why the phase separations should start small and increase, and why the mixing times should decrease, but the general interference process remains consistent throughout the iterations.

\section{A Few Important Mixers}
\label{sec:mixers}
This section addresses the mixing unitary and its efficient implementation for three cases: binary problems ($k=2$), integer problems ($k>2$) and permutation problems ($k=n$, no repetition). The action of each of these mixers on an arbitrary solution state can be shown to satisfy the necessary condition in \eqref{eq:mixer_requirement_1}.

\subsection{Binary Mixer (Hypercube Graph)}
Consider a combinatorial problem consisting of $n$ binary decision variables ${x_j \in \{0,1\}}$, where solutions are characterised by vectors ${\bm{x}=(x_1,x_2,...,x_n)}$ and the space of feasible solutions $S$ contains all such vectors, such that $N=2^n$. Each binary variable is naturally encoded in the computational basis state of a single qubit (where the binary and one-hot encodings are equivalent), so that the solution states are the computational basis states of an $n$ qubit register. The equal superposition is efficiently prepared,
\begin{equation}
    \ket{s} = H^{\otimes n} \ket{0}^{\otimes n} = \frac{1}{\sqrt{2^n}} \sum_{\bm{x} \in S} \ket{\bm{x}}.
\end{equation}

The set of $n$ bit-flips generate nearest neighbour solutions, so the set of $n$ Pauli X operators $\sigma_x$ generate nearest neighbour solution states. The adjacency matrix can therefore be expressed as, 
\begin{equation}
    A = \sum_{j=1}^{n}  \id^{\otimes j-1} \otimes \sigma_x \otimes \id^{\otimes n-j},
    \label{eq:A_binary}
\end{equation}
which is the adjacency matrix for an n-dimensional hypercube. The continuous-time quantum walk on the hypercube, and hence the mixing unitary, can be expressed as, 
\begin{equation}
    U_M(t) = e^{-\text{i} t A} = \prod_{j=1}^n e^{-\text{i} t \sigma_x},
\end{equation}
which is identical to the transverse-field Hamiltonian from the QAOA, and is efficiently implementable with single qubit rotations.

\subsection{Integer Mixer (Hamming Graph)}
Consider a combinatorial problem consisting of $n$ integer decision variables ${x_j \in \{0,1,...,k-1\}}$, where solutions are characterised by vectors ${\bm{x}=(x_1,x_2,...,x_n)}$ and the space of feasible solutions $S$ contains all such vectors, such that ${N=k^n}$. Each integer variable can be encoded in the computational basis state of a sub-register of $m=\lceil \log_2 k \rceil$ qubits for a binary encoding or $m=k$ qubits for a one-hot encoding. The equal superposition state is efficiently prepared over an $n m$ qubit register containing $n$ sub-registers,
\begin{equation}
    \ket{s} = \left(U_k \ket{0}^{\otimes m} \right)^{\otimes n} = \ket{k}^{\otimes n} = \frac{1}{\sqrt{k^n}} \sum_{\bm{x} \in S} \ket{\bm{x}},
\end{equation}
where $\ket{k}$ is the equal superposition over the $k$ utilized computational basis states of a sub-register (those which directly encode each of the possible decision variable values). Each circuit implementation (one-hot and binary) of the unitary $U_k$ which prepares $\ket{k}$ is described in \cite{longPaper}.

When modifying a particular solution to return a nearest neighbour, any one of the decision variables can be modified, and there are $(k-1)$ values which could replace each of them. As such, there are $n(k-1)$ moves which generate the nearest neighbours. The adjacency matrix of the mixing graph which connects these nearest neighbour solution states can be expressed as, 
\begin{equation}
    A = \sum_{j=1}^n \id^{\otimes j-1}  \otimes K_k  \otimes \id^{\otimes n-j},
    \label{eq:A_integer}
\end{equation}
where $K_k$ is the adjacency matrix of a complete graph connecting the $k$ utilized computational basis states within a sub-register, ${K_k = k \ketbra{k}{k} - \id}$. This mixing graph (the subgraph connecting feasible solutions) is also known as a Hamming graph $H(n,k)$, and is the Cartesian product of $n$ complete graphs $K_k$.

As the individual terms in the adjacency matrix are commuting, the mixer can be implemented by applying a CTQW on the complete graph $K_k$ within each sub-register,
\begin{equation}
    U_M(t) = e^{-\text{i} n t} e^{-\text{i} t A} = e^{-\text{i} n t} \prod_{j=1}^n e^{-\text{i} t K_k},
\end{equation}
where a global phase has been introduced in order that the expression simplifies,
\begin{equation}
    U_M(t) = \prod_{j=1}^n \left(  e^{-\text{i} k t} \ketbra{k}{k}  + \left( \id - \ketbra{k}{k} \right) \right).
\end{equation}

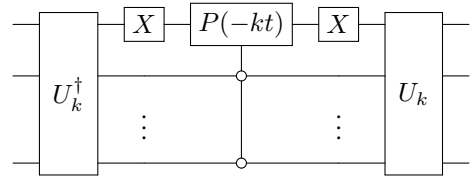
\begin{figure}[htbp]
    \centering
    \[ \Qcircuit @C=1em  @R=0.7em  {
        & \multigate{3}{U_k^\dagger} & \gate{X} & \gate{P(-k t)} & \gate{X} & \multigate{3}{U_k} & \qw \\
        & \ghost{U_k^\dagger} & \qw & \ctrlo{-1} & \qw & \ghost{U_k} & \qw \\
        & \push{\rule{0em}{1em}} & \colorbox{white}{\vdots} & & \colorbox{white}{\vdots} & \\
        & \ghost{U_k^\dagger} & \qw & \ctrlo{-2} & \qw & \ghost{U_k} & \qw \\
    }\]
    \caption{Quantum circuit implementation of the CTQW on a complete graph (${e^{-\text{i} k t} \ketbra{k}{k}  + \left( \id - \ketbra{k}{k} \right)}$) to be applied within each of the $m$-qubit sub-registers. $P(\phi)$ is the phase gate which maps computational basis states ${\ket{0} \mapsto \ket{0}}$ and $\ket{1} \mapsto e^{\text{i} \phi}\ket{1}$.}
    \label{fig:circuit}
\end{figure}

A quantum circuit implementation of the mixer can be achieved by applying the circuit in Fig.~\ref{fig:circuit} within each of the sub-registers.

\subsection{Permutation Mixer (Transposition Graph)}
Consider a combinatorial problem consisting of $n$ integer decision variables ${x_j \in \{0,1,...,n-1\}}$, where solutions are characterised by vectors ${\bm{x}=(x_1,x_2,...,x_n)}$ and the space of feasible solutions $S$ contains all such vectors without repeating elements, such that $N=n!$. As with the integer problems, each integer variable can be encoded in the computational basis state of a sub-register of $m=\lceil \log_2 n \rceil$ or $m=k$ qubits. A recursive method for generating the equal superposition over feasible solution states is presented in \cite{longPaper}, which has gate complexity of order $n^3$, and works for both the binary and one-hot encodings. An alternative method is presented in \cite{Grover_mixer}, effective for just the one-hot encoding.

The set of $\frac{1}{2} n(n-1)$ possible transpositions (the swapping of two elements of the vector $\bm{x}$) generates nearest neighbour solutions. As such, the set of all possible sub-register SWAP operations generates nearest neighbour solution states, so the adjacency matrix of the mixing graph can be expressed as,
\begin{equation}
    A = \sum_{i=0}^{n-2} \sum_{j=i+1}^{n-1} \text{SWAP}_{i,j},
\end{equation}
where $\text{SWAP}_{i,j}$ is defined as the permutation matrix associated with swapping the states of the $i^\text{th}$ and $j^\text{th}$ sub-registers (applying identity to the remaining sub-registers). This mixing graph (the subgraph connecting feasible solutions) is a transposition graph.

Unlike the terms in \eqref{eq:A_binary} and \eqref{eq:A_integer}, the individual $\text{SWAP}_{i,j}$ operators composing this adjacency matrix do not commute. As such, the application of this mixer is not as trivial. However it can be efficiently implemented via Hamiltonian simulation. For example, Berry \emph{et al.} \cite{berry2015simulating} present a truncated Taylor series method to implement $e^{-\text{i} H t}$ where $H$ is a linear combination of implementable unitary matrices, which is exactly satisfied in this case with $H=A$ and $\text{SWAP}_{i,j}$ playing the role of the individual unitary matrices which are efficiently implemented with regular 2-qubit SWAP gates. 

\section{Simulation Results}
\label{sec:results}

This section introduces several problems and provides simulation results for the non-variational QWOA applied to each of them. For this purpose, a single random problem instance has been generated for each, with the details provided in \cite{longPaper}. In addition, each of the generated problems are shown to satisfy the necessary condition in \eqref{eq:mixer_requirement_2}, where the result is derived analytically for weighted maxcut, and is demonstrated via statistical sampling for the other problems. 

During physical implementation, derivative-free optimisation methods such as Bayesian optimisation or CMA-ES, are likely to be more effective than gradient-based methods, as they are robust to uncertainties in the approximated values of the function to be optimised. This allows them to be effective with limited shots (repeated state-preparation and measurements). However, for the following results, parameter values have all been determined via the BFGS algorithm, to optimise for expectation value of $f(\bm{x})$, initialised with $\gamma=1$, $t=0.1$ and $\beta=\frac{1}{p}$ (and $\bm{\lambda}_T=\bm{\lambda}_F$ where relevant).

Note that the approximation ratio, defined as,
\begin{equation}
    \text{Approximation Ratio} = \frac{f(\bm{x})}{\text{optimal}\{f(\bm{x}):\bm{x} \in S\}},
\end{equation}
is used for some of the figures, providing clarity, as regardless of the problem or problem instance, an optimal solution has approximation ratio equal to $1$.

\subsection{Weighted Maxcut}
\begin{figure}[htbp]
    \centering
    \includegraphics[width=0.95\columnwidth]{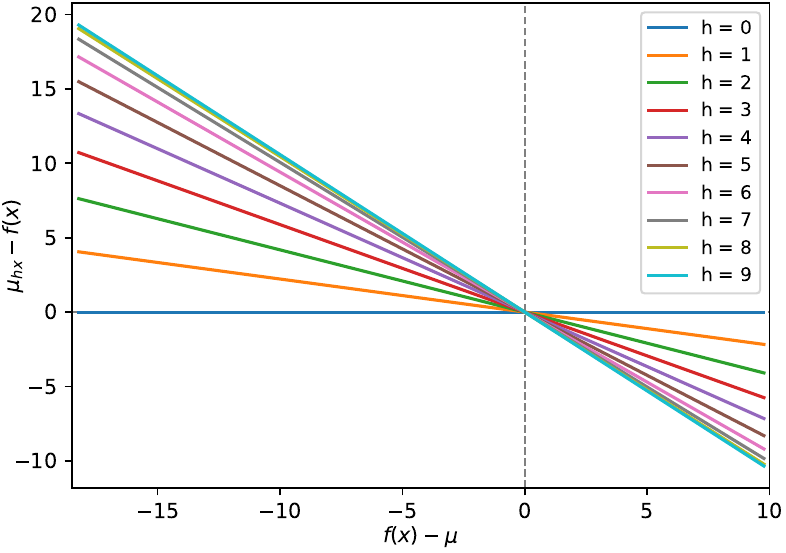}
    \vspace{-0.2cm}
    \caption{Maxcut: An analysis of objective function values, as distributed over the binary/hypercube mixing graph.}
    \label{fig:maxcut_subset_means}
\end{figure}

A weighted maxcut problem of size $n$ is characterised by an $n$ vertex weighted graph $G(V,E,W)$, where the aim is to find a partitioning of the graph's vertices into two subsets such that the total weight of edges passing between the two subsets is maximised. As such, solutions to the problem can be characterised as a vector of binary decision variables, where each decision variable assigns a vertex to one subset or the other. The objective function can then be expressed,
\begin{equation}
    f(\bm{x}) = \sum_{(i,j) \in E} w_{ij} (x_i - x_j)^2,
\end{equation}
which satisfies the condition in \eqref{eq:mixer_requirement_2} when solutions are distributed over the binary mixing graph (hypercube), as illustrated in Fig.~\ref{fig:maxcut_subset_means}.

\begin{figure}[htbp]
    \centering
    \includegraphics[width=0.95\columnwidth]{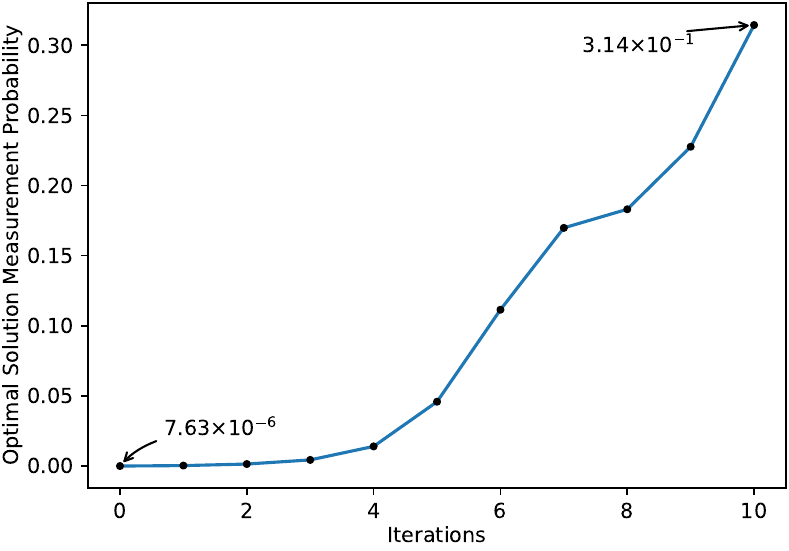}
    \vspace{-0.2cm}
    \caption{Maxcut: Measurement probability of the optimal solution throughout the $10$ iterations preparing the amplified state $\ket{\gamma, t}$.}
    \label{fig:maxcut_opt_probs}
\end{figure}

\begin{figure}[htbp]
    \centering
    \includegraphics[width=0.95\columnwidth]{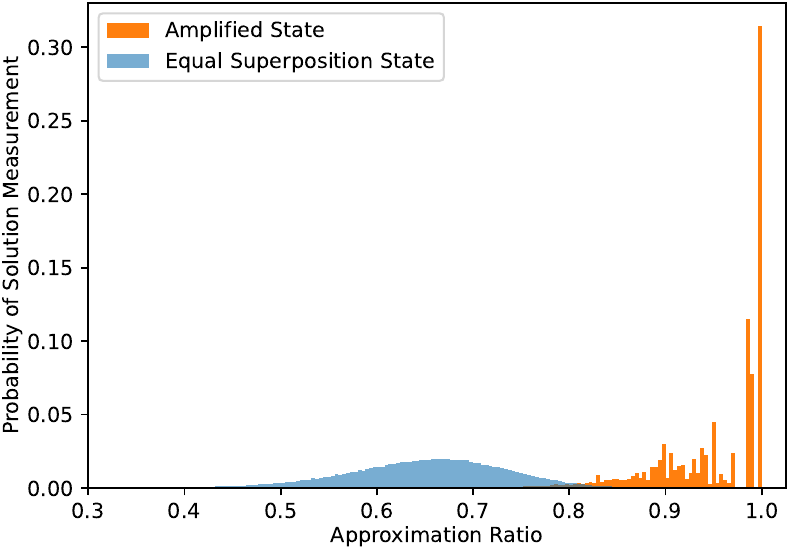}
    \vspace{-0.2cm}
    \caption{Maxcut: Distributions for approximation ratio as measured from the initial equal superposition state, $\ket{s}$, and the amplified state $\ket{\gamma, t}$.}
    \label{fig:maxcut_distributions}
\end{figure}

With the binary mixer and $p=10$, the optimal amplified state $\ket{\gamma,t,\beta}$ for an $n=18$ problem instance is given by $\gamma=2.4340$, $t=0.4517$ and $\beta=0.2844$ and is analysed in Fig.~\ref{fig:maxcut_opt_probs} and Fig.~\ref{fig:maxcut_distributions}, demonstrating that the non-variational QWOA solves this particular instance.

\subsection{Maximum Independent Set}
\begin{figure}[htbp]
    \centering
    \includegraphics[width=0.95\columnwidth]{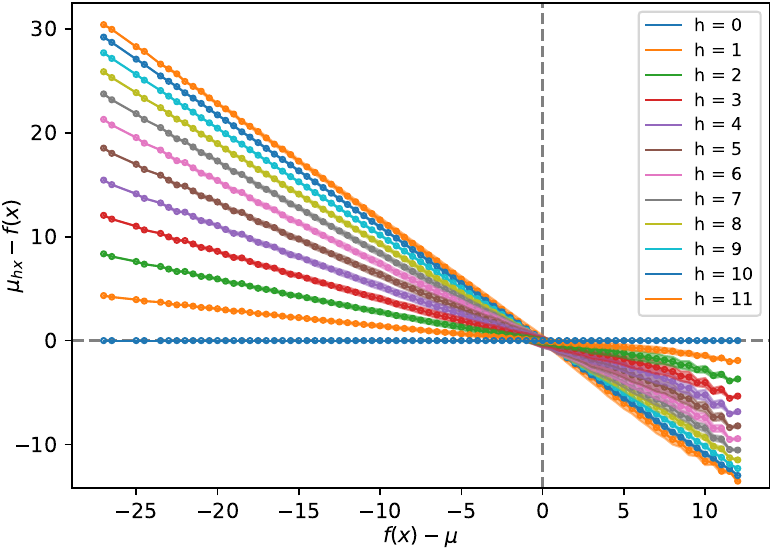}
    \vspace{-0.2cm}
    \caption{Maximum independent set: An analysis of objective function values with $\bm{\lambda}_F = \{1.5,0\}$, as distributed over the binary/hypercube mixing graph. The points and shading show the mean and $\pm$ a single standard deviation from 200 equally spaced bins (where the bins were not empty).}
    \label{fig:MIS_subset_means}
\end{figure}

A maximum independent set problem of size $n$ is characterised by an $n$ vertex graph $G(V,E)$, where the aim is to find one or all of the subsets of the graph's vertices which include the maximum number of vertices without including any pairs of connected vertices. Solutions to the problem can be characterised by a length $n$ vector of binary decision variables, one for each of the vertices of the graph, encoding whether or not each vertex is included in the subset. The penalised objective function (to be maximised) can then be expressed,
\begin{equation}
    f(\bm{x})_{\bm{\lambda}} = \left( \sum_{j=1}^n x_j \right) - \lambda_1 P_1(\bm{x}) - \lambda_2 P_2(\bm{x}),
\end{equation}
where $P_1$ counts the number of connected pairs in the subset, 
\begin{equation}
    P_1(\bm{x}) = \sum_{(i,j) \in E} x_i x_j,
\end{equation}
and $P_2(\bm{x})$ flags whether the solution violates the constraint,
\begin{equation}
    P_2(\bm{x}) = \begin{cases}
        0,\hspace{0.5cm} \mbox{if $ \sum_{(i,j) \in E} x_i x_j = 0$} \\
        \vspace{-0.25cm} \\
        1, \hspace{0.5cm} \mbox{otherwise} \\
    \end{cases}.
\end{equation}
Fig.~\ref{fig:MIS_subset_means} shows that the objective function satisfies the necessary condition in \eqref{eq:mixer_requirement_2}, for solutions distributed over the binary mixing graph.

\begin{figure}[htbp]
    \centering
    \includegraphics[width=0.9\columnwidth]{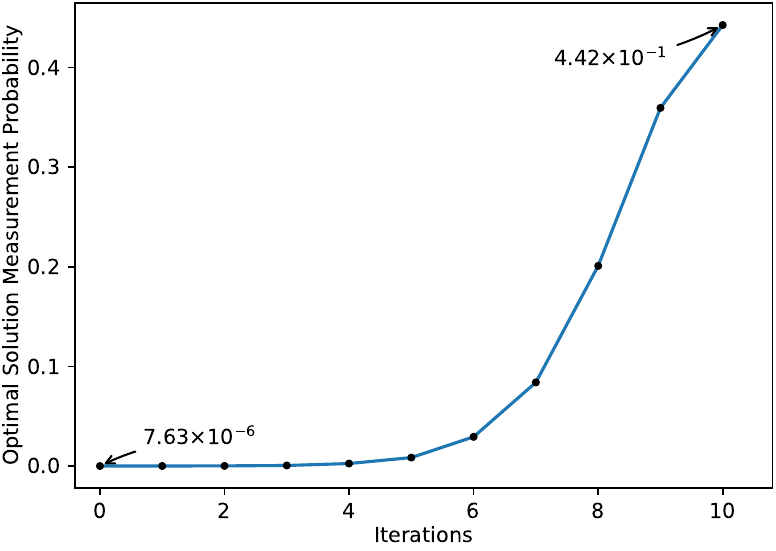}
    \vspace{-0.2cm}
    \caption{Maximum independent set: Measurement probability of the two optimal solutions throughout the $10$ iterations preparing the amplified state $\ket{\gamma,t,\beta}_{\bm{\lambda}_T}$.}
    \label{fig:MIS_opt_probs}
\end{figure}

\begin{figure}[htbp]
    \centering
    \includegraphics[width=0.95\columnwidth]{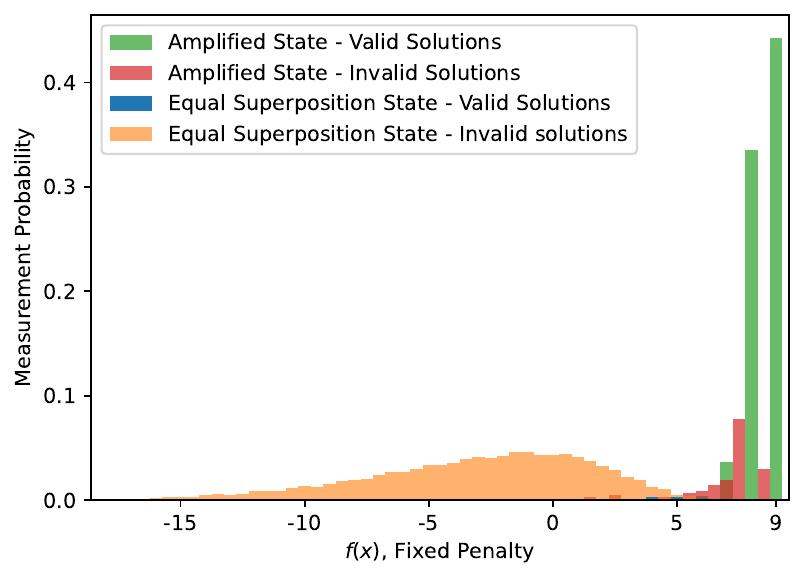}
    \vspace{-0.2cm}
    \caption{Maximum independent set: Distributions for $f(\bm{x})_{\bm{\lambda}_F}$ as measured from the initial equal superposition state, $\ket{s}$, and the amplified state $\ket{\gamma,t,\beta}_{\bm{\lambda}_T}$.}
    \label{fig:MIS_distributions}
\end{figure}

With the binary mixer, $p=10$ and $\bm{\lambda}_F = \{1.5,0\}$, the optimal amplified state  $\ket{\gamma,t,\beta}_{\bm{\lambda}_T}$ for an $n=18$ problem instance is given by parameter values, $\gamma=3.0098$, $t=0.5724$, $\beta=0.1722$ and ${\bm{\lambda}_T=(1.0370,0.5235)}$ and is analysed in Fig.~\ref{fig:MIS_opt_probs} and Fig.~\ref{fig:MIS_distributions}, demonstrating that the non-variational QWOA solves this particular instance, preferentially amplifying the measurement probability of the two maximum independent sets.

In a recent study focused on quantum optimisation for constrained problems, Saleem \emph{et al.} \cite{saleem2023approaches} focus specifically on the maximum independent set problem and introduce an alternative variational approach, which was demonstrated on a randomly generated 14 vertex graph. While it is difficult to make direct comparison, we note that the non-variational QWOA solves this particular problem instance, producing a measurement probability for the 8 maximum independent sets of $0.16$ for $p=2$ iterations, and $0.36$ for $p=3$ iterations.

\subsection{k-means Clustering}
Cluster analysis involves dividing a set of data-points into clusters based on similarity. k-means clustering measures the quality of a clustering by the degree to which it minimises the sum of squared euclidean distance between each data-point and its respective cluster's centroid. That is, k-means clustering seeks a partitioning of $n$ multi-dimensional real vectors (data-points) $\bm{v}_j$ into $k$ clusters which minimises the objective function,
\begin{equation*}
    f(\bm{x}) = \sum_{i=0}^{k-1} \frac{1}{\abs{C_i(\bm{x})}} \sum_{\bm{a},\bm{b}\in C_i(\bm{x})} \norm{\bm{a}-\bm{b}}^2,
\end{equation*}
where $C_i(\bm{x}) = \{\bm{v}_j : x_j=i\}$ is cluster $i$ and $\norm{\bm{a}-\bm{b}}$ is the euclidean distance between data points $\bm{a}$ and $\bm{b}$ which are contained in $C_i$.

\begin{figure}[htbp]
    \centering
    \includegraphics[width=0.95\columnwidth]{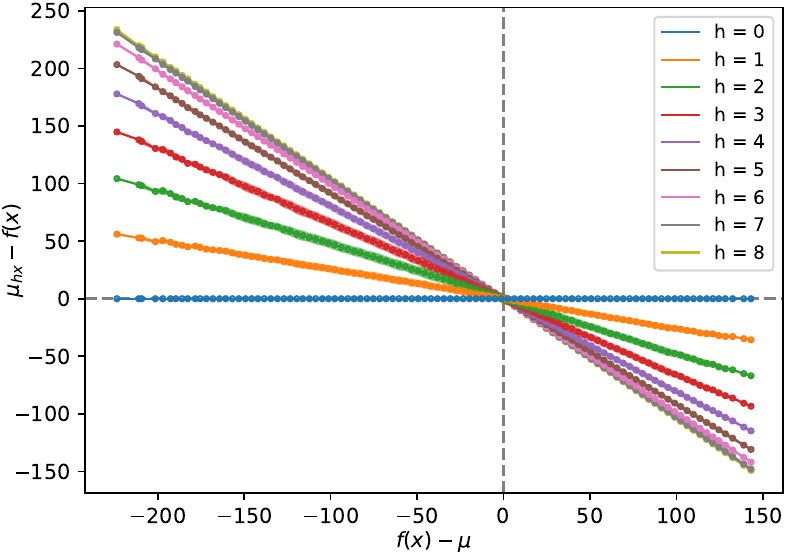}
    \vspace{-0.2cm}
    \caption{k-means clustering: An analysis of objective function values, as distributed over the integer mixing graph. The points and shading show the mean and $\pm$ a single standard deviation from 200 equally spaced bins (where the bins were not empty).}
    \label{fig:kmeans_subset_means}
\end{figure}

An additional adjustment is made to the objective function to improve its adherence to the condition in \eqref{eq:mixer_requirement_2},
\begin{equation*}
    f(\bm{x}) \xrightarrow{} f(\bm{x}) - (\mu_{c(\bm{x})}-\mu_k),
\end{equation*}
where $c(\bm{x})$ is the number of non-empty clusters in solution $\bm{x}$, and $\mu_j$ is defined as the mean objective function value of solutions with $j$ non-empty clusters. As can be seen in Fig.~\ref{fig:kmeans_subset_means}, this modified objective function, for solutions distributed over the vertices of the integer mixing graph, conforms well with the necessary condition in \eqref{eq:mixer_requirement_2}.

\begin{figure}[htbp]
    \centering
    \includegraphics[width=0.95\columnwidth]{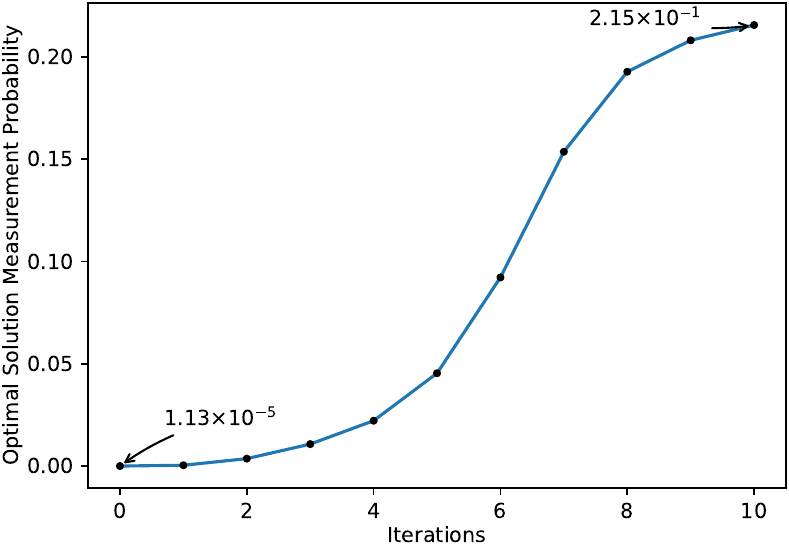}
    \vspace{-0.2cm}
    \caption{k-means clustering: Measurement probability of the optimal solution throughout the $10$ iterations preparing the amplified state $\ket{\gamma, t}$.}
    \label{fig:kmeans_opt_probs}
\end{figure}

With the integer mixer and $p=10$, the optimal amplified state $\ket{\gamma,t,\beta}$ for an $n=12$, $k=3$ problem instance is given by parameter values $\gamma=1.5345$, $t=0.2483$ and $\beta=0.3441$ and is analysed in Fig.~\ref{fig:kmeans_opt_probs}, demonstrating that the non-variational QWOA solves this particular instance.

\subsection{The Capacitated Facility Location Problem}
\label{sec:CFLP}
The facility location problem involves the servicing of $n$ customers via some number of facilities located amongst $k$ candidate locations. A solution to the problem is a selection of candidate locations at which to install a facility, and the assignment of each customer to a facility. As such each solution to the problem can be characterised by a vector, ${\bm{x} = (x_1, x_2, ..., x_n)}$, where ${x_j \in \{0,1,...,k-1\}}$ specifies the allocation of customer $j$ to a facility at candidate location $x_j$.

There are a few parameters which define a particular problem instance. $F_i$ is the cost associated with opening a facility at candidate location $i$, and $L_{j,i}$ measures the transport cost between customer $j$ and facility location $i$. We consider a variant of the problem where each customer is serviced by only a single facility. In addition, we consider the capacitated variant, where customer $j$ requires a number of resources $R_j$, and a facility at candidate location $i$ has a maximum capacity $C_i$ with regards to total supplied resources.  

Prior to embedding the capacity constraint via penalty terms, the objective function to be minimised is given by,
\begin{equation}
    f(\bm{x}) =  \sum_{j=1}^n R_j L_{j,x_j}  + \sum_{i = 0}^{k-1} \begin{cases}
        F_i,\hspace{0.5cm} \mbox{if $i \in \bm{x}$} \\
        \vspace{-0.25cm} \\
        0, \hspace{0.5cm} \mbox{otherwise} \\
        \end{cases},
    \label{eq:non_penalised}
\end{equation}
where the first sum accounts for the transportation cost of resources and the second sum accounts for the total cost of opening facilities. Introducing facility capacities, the penalised objective function can be expressed as,
\begin{equation}
    \label{eq:lambda3}
    f(\bm{x})_{\bm{\lambda}} = g(\bm{x})_{\bm{\lambda}} - \begin{cases}
        0,\hspace{0.5cm} \mbox{if $\bm{x}$ is valid} \\
        \vspace{-0.25cm} \\
        \lambda_3 \left[ g(\bm{x})_{\bm{\lambda}} - g(\bm{y})_{\bm{\lambda}} \right], \hspace{0.1cm} \mbox{otherwise} \\
    \end{cases}
\end{equation}
where $\bm{y}$ is the solution (or at least an approximate solution) to the unconstrained variant and,
\begin{equation}
    g(\bm{x})_{\bm{\lambda}} = f(\bm{x}) + \sum_{i = 1}^k \begin{cases}
        0,\hspace{0.5cm} \mbox{if $ \sum_{j: x_j = i} R_j \leq C_i$} \\
        \vspace{-0.25cm} \\
        \lambda_1 P_{i,1}(\bm{x}) + \lambda_2 P_{i,2}(\bm{x}), \hspace{0.1cm} \mbox{otherwise} \\
    \end{cases}.
\end{equation}

\begin{figure}[htbp]
    \centering
    \includegraphics[width=0.95\columnwidth]{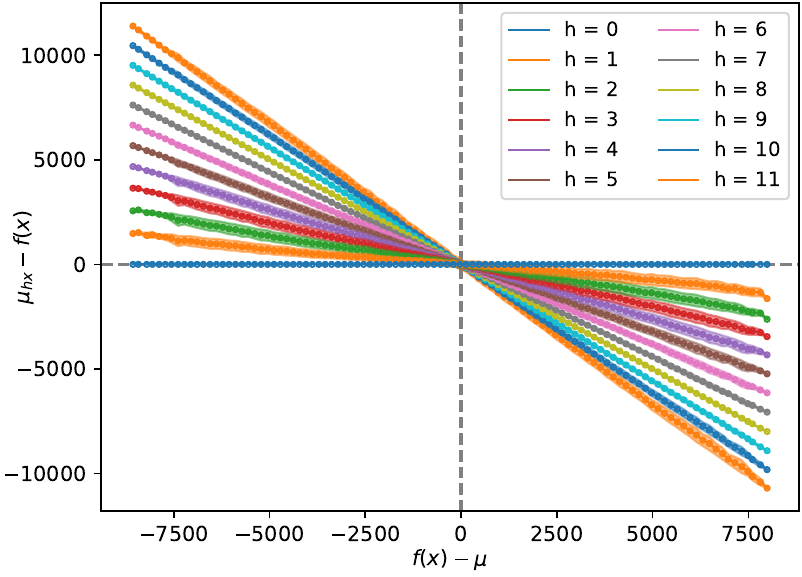}
    \vspace{-0.2cm}
    \caption{Capacitated facility location problem: An analysis of objective function values $f(\bm{x})_{\bm{\lambda}_T}$ as distributed over the integer mixing graph. The points and shading show the mean and $\pm$ a single standard deviation from 200 equally spaced bins (where the bins were not empty).}
    \label{fig:CFLP_subset_means}
\end{figure}

\noindent $P_{i,1}(\bm{x})$ and $\lambda_1$ apply a penalty when the capacity of facility $i$ is exceeded,
\begin{equation}
    P_{i,1}(\bm{x}) = \left(\frac{1}{k} \sum_{j=1}^k F_j\right)  \left\lceil \frac{\left( \sum_{j: x_j = i} R_j \right) - C_i}{C_i} \right\rceil ,
\end{equation}
which is proportional to the average facility opening cost. $P_{i,2}(\bm{x})$ and $\lambda_2$ provide a penalty proportional to the number of excess resources at facility $i$, 
\begin{equation}
    P_{i,2}(\bm{x}) = \left(\frac{1}{n k} \sum_{j=1}^n\sum_{l=1}^k L_{j,l}\right)  \left( \left( \sum_{j: x_j = i} R_j \right) - C_i\right),
\end{equation}
which scales with the average facility-customer distance. The purpose of \eqref{eq:lambda3} and hence the role of $\lambda_3$ is to correct for bimodality which is introduced by the penalties in $g(\bm{x})_{\bm{\lambda}}$. Fig.~\ref{fig:CFLP_subset_means} shows that the penalised objective function conforms well with the condition in \eqref{eq:mixer_requirement_2}, which is generated for a random $n=12$, $k=3$ problem and is shown with the tuned penalty parameters. 

\begin{figure}[htbp]
    \centering
    \includegraphics[width=0.92\columnwidth]{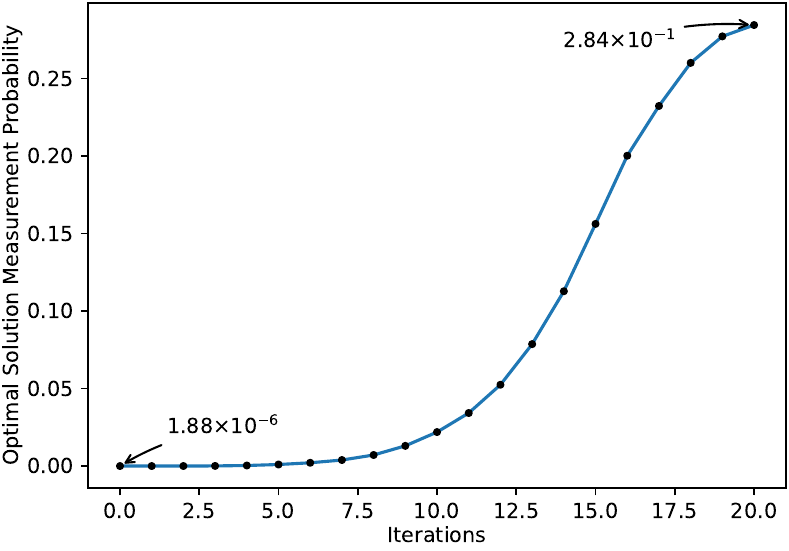}
    \vspace{-0.2cm}
    \caption{Capacitated facility location problem: Measurement probability of the optimal solution throughout the $20$ iterations preparing the amplified state $\ket{\gamma, t}_{\bm{\lambda}_T}$.}
    \label{fig:CFLP_opt_probs}
\end{figure}

\begin{figure}[htbp]
    \centering
    \includegraphics[width=0.95\columnwidth]{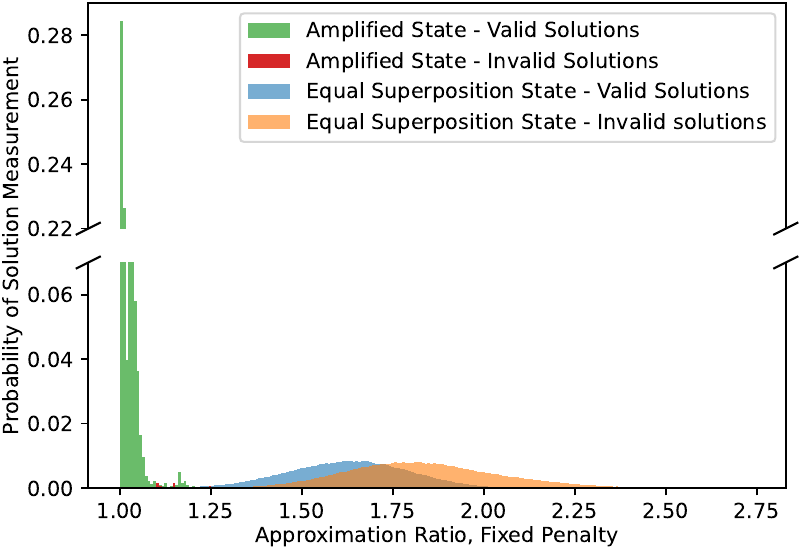}
    \vspace{-0.2cm}
    \caption{Capacitated facility location problem: Distributions for $f(\bm{x})_{\bm{\lambda}_F}$ as measured from the initial equal superposition state, $\ket{s}$, and the amplified state $\ket{\gamma,t,\beta}_{\bm{\lambda}_T}$.}
    \label{fig:CFLP_distributions}
\end{figure}

With the integer mixer, $p=20$ and ${\bm{\lambda}_F=(1,1,0)}$, the optimal amplified state $\ket{\gamma,t,\beta}_{\bm{\lambda}_T}$ for an $n=12$, $k=3$ problem instance is given by $\gamma=2.5732$, $t=0.2756$, $\beta=0.0593$ and ${\bm{\lambda}_T=(0.8966,0.4996,0.1732)}$ and is analysed in Fig.~\ref{fig:CFLP_opt_probs} and Fig.~\ref{fig:CFLP_distributions}, demonstrating that the non-variational QWOA solves this particular instance.

\subsection{The Quadratic Assignment Problem}
\label{sec:QAP}
The quadratic assignment problem (QAP) is an extremely challenging combinatorial optimisation problem \cite{QAP_hardness_1} which makes the problem an interesting candidate for the study of quantum algorithms. Consider some set of facilities between which materials must be transported. A QAP can be characterised as the problem of assigning each of these facilities to a location, where the number of candidate locations is equal to the number of facilities. Each problem instance can be fully characterised by the distances (or costs associated with transport) between candidate locations and the amount of material flowing between facilities. We define $L_{i,j}$ as the distance or transport-cost between location $i$ and location $j$. Likewise, $F_{i,j}$ quantifies the material flowing between facility $i$ and facility $j$. Consider a solution to a QAP of size $n$ to be characterised by a vector, ${\bm{x} = (x_0, x_1, ..., x_{n-1})}$, where ${x_j \in \{0,1,...,n-1\}}$ specifies the allocation of a facility $j$ to location $x_j$. Only one facility can be assigned to each location, so we define the space of valid solutions $S$ to contain all $N=n!$ permutations, or in other words, all $\bm{x}$ such that there is no repetition in $x_j$. Given a particular solution $\bm{x}$, the objective function (to be minimised) can be expressed as,
\begin{equation}
    f(\bm{x}) = \sum_{i,j} F_{i,j} L_{x_i,x_j}.
    \label{eq:obj_func_QAP}
\end{equation}

\begin{figure}[htbp]
    \centering
    \includegraphics[width=0.95\columnwidth]{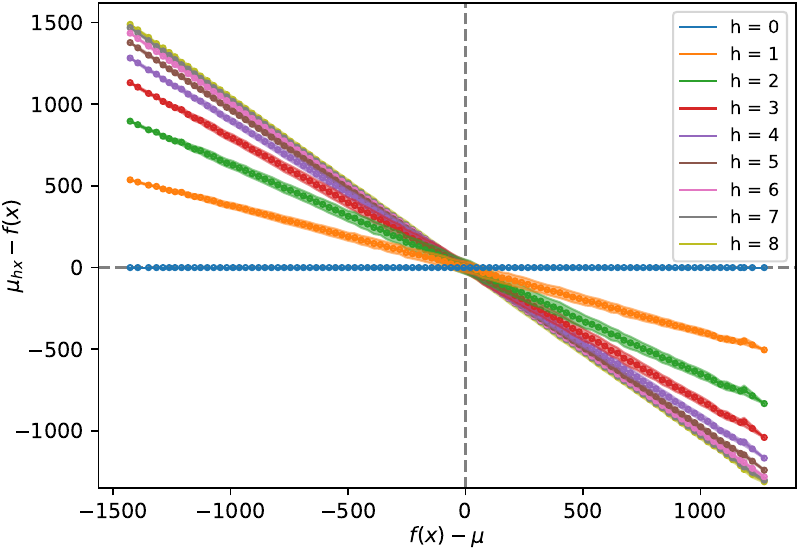}
    \vspace{-0.2cm}
    \caption{The quadratic assignment problem: An analysis of objective function values $f(\bm{x})$ as distributed over the permutation mixing graph. The points and shading show the mean and $\pm$ a single standard deviation from 200 equally spaced bins (where the bins were not empty).}
    \label{fig:QAP_subset_means}
\end{figure}

The objective function for the randomly generated $n=9$ problem instance satisfies the condition in \eqref{eq:mixer_requirement_2} when solutions are distributed over the permutation mixing graph, as illustrated in Fig.~\ref{fig:QAP_subset_means}.

\begin{figure}[htbp]
    \centering
    \includegraphics[width=0.95\columnwidth]{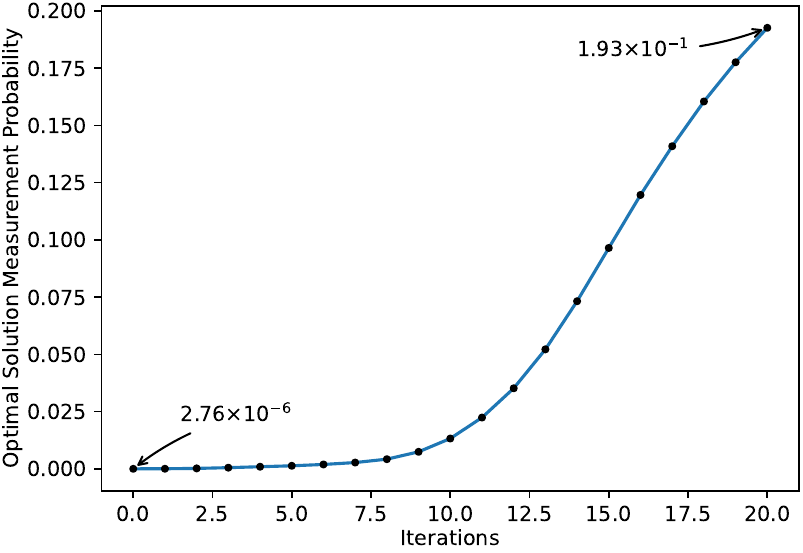}
    \vspace{-0.2cm}
    \caption{The quadratic assignment problem: Measurement probability of the optimal solution throughout the $20$ iterations preparing the amplified state $\ket{\gamma, t}$.}
    \label{fig:QAP_opt_probs}
\end{figure}

With the permutation mixer and $p=20$, the optimal amplified state $\ket{\gamma,t,\beta}$ for an $n=9$ problem instance is given by $\gamma=1.2636$, $t=0.1219$ and $\beta=0.4167$ and is analysed in Fig.~\ref{fig:QAP_opt_probs}, demonstrating that the non-variational QWOA solves this particular instance.

\section{Conclusion}
This paper presents a strong case for algorithm development motivated from first principles. The typical variational approach for quantum combinatorial optimisation offloads much of the heavy lifting to parameterised black boxes, whose inner workings we often don't understand. By interpreting the mixing unitary as a continuous time quantum walk and analysing the statistics of objective function values distributed over a mixing graph, we can understand the interference process produced via the alternating application of a phase-separation and a mixing unitary. This theoretical framework is at the heart of the non-variational QWOA and enables its generalisation and application to a wide range of practically important and intractable optimisation problems, as we have demonstrated. 

We characterise the algorithm as non-variational because, given a number of iterations $p$, the amplified state is determined by a small set of parameters $\{\sigma, \beta, \gamma,t\}$. The number of iterations necessary for a particular problem and problem size can be determined via benchmarking, or may be limited by achievable circuit depths for a particular quantum computing device. The approximate standard deviation $\sigma$ of $f(\bm{x})$ can be easily determined via small scale random sampling. The user-specified parameters $\gamma$, $t$ and $\beta$, produce significant amplification of optimal and near-optimal solutions across a wide range of values. In addition, the optimal values are relatively consistent across problem instances, and can also be found (or improved) via 3-dimensional local optimisation. 

The number of state-preparation and measurements required to perform this local optimisation is fixed, independent of iteration count or problem size. Hence, one important component of future work will involve the deployment of gradient-free optimisation methods, such as Bayesian optimisation and CMA-ES, in order to demonstrate the ability to find/improve parameter values within a fixed and small number of total state preparations/measurements ($<10000$).

In order to quantify the algorithm's average-case performance, and demonstrate advantage relative to classical heuristics, such as local search based methods, another important future area of research will involve detailed and large scale benchmarking for various problems. 

We also plan to demonstrate that the use of CTQWs on problem specific graphs provides significant advantage for non-binary problems, such as the quadratic assignment problem, over the alternative approach of converting these to QUBO problems and making use of the standard QAOA/binary mixer.

\bibliographystyle{ieeetr}
\bibliography{refs}

\end{document}